\documentclass[conference]{IEEEtran}
\IEEEoverridecommandlockouts

\usepackage{cite}
\usepackage{amsmath,amssymb,amsfonts}
\usepackage{algorithmic}
\usepackage{graphicx}
\usepackage{textcomp}
\usepackage{xcolor}
\usepackage{booktabs}
\usepackage{multirow}
\usepackage{hyperref}
\usepackage{balance}
\usepackage{subcaption}

\def\BibTeX{{\rm B\kern-.05em{\sc i\kern-.025em b}\kern-.08em
    T\kern-.1667em\lower.7ex\hbox{E}\kern-.125emX}}

\begin{document}

\title{Context-Aware Phishing Email Detection Using Machine Learning and NLP}
\makeatletter
\newcommand{\linebreakand}{%
  \end{@IEEEauthorhalign}
  \hfill\mbox{}\par
  \mbox{}\hfill\begin{@IEEEauthorhalign}
}
\makeatother

\author{
\IEEEauthorblockN{Amitabh Chakravorty}
\IEEEauthorblockA{
School of Information Technology\\
University of Cincinnati\\
Cincinnati, OH, USA\\
chakraa4@mail.uc.edu
}
\and
\IEEEauthorblockN{Matthew Price}
\IEEEauthorblockA{
School of Information Technology\\
University of Cincinnati\\
Cincinnati, OH, USA\\
pricem6@mail.uc.edu
}
\and
\IEEEauthorblockN{Nelly Elsayed}
\IEEEauthorblockA{
School of Information Technology\\
University of Cincinnati\\
Cincinnati, OH, USA\\
elsayeny@ucmail.uc.edu
}
\linebreakand
\IEEEauthorblockN{Zag ElSayed}
\IEEEauthorblockA{
School of Information Technology\\
University of Cincinnati\\
Cincinnati, OH, USA\\
elsayezs@ucmail.uc.edu
}
}
\maketitle

\begin{abstract}
Phishing attacks remain among the most prevalent cybersecurity threats, causing significant financial losses for individuals and organizations worldwide. This paper presents a machine learning-based phishing email detection system that analyzes email body content using natural language processing (NLP) techniques. Unlike existing approaches that primarily focus on URL analysis, our system classifies emails by extracting contextual features from the entire email content. We evaluated two classification models, Na\"ive Bayes and Logistic Regression, trained on a combined corpus of 53,973 labeled emails from three distinct datasets. Our preprocessing pipeline incorporates lowercasing, tokenization, stop-word removal, and lemmatization, followed by Term Frequency-Inverse Document Frequency (TF-IDF) feature extraction with unigrams and bigrams. Experimental results demonstrate that Logistic Regression achieves 95.41\% accuracy with an F1-score of 94.33\%, outperforming Na\"ive Bayes by 1.55 percentage points. The system was deployed as a web application with a FastAPI backend, providing real-time phishing classification with average response times of 127ms.
\end{abstract}
\begin{IEEEkeywords}
Phishing detection, machine learning, natural language processing, email security, TF-IDF, logistic regression, text classification
\end{IEEEkeywords}
\section{Introduction}

Phishing remains one of the most significant threats to digital security due to its simplicity and high return on investment for attackers \cite{moura2024}. Modern phishing campaigns have evolved beyond simple deceptive emails to sophisticated social engineering attacks that leverage artificial intelligence for content generation \cite{olea2025}. Recent studies indicate that approximately 66 million business email compromise attacks are detected monthly, demonstrating the sheer volume of malicious communications targeting organizations \cite{weinz2025}. 

According to the Cybersecurity and Infrastructure Security Agency (CISA), eight out of ten organizations reported that at least one employee had fallen victim to a phishing email. Moreover, within the first ten minutes of receiving a malicious message, 84\% of recipients interacted with the email by replying, clicking a spoofed link, or opening a malicious attachment~\cite{cisa_phishing_2022}. These statistics highlight the critical need for automated, accurate phishing-detection mechanisms that operate in real time.

A core challenge in phishing detection lies in the constantly evolving tactics used by attackers. Traditional rule-based systems struggle to adapt to new attack vectors and require frequent manual updates, while purely URL-based detection methods become ineffective when attackers utilize compromised legitimate websites. These limitations motivate the development of content-aware detection systems that analyze the linguistic and semantic characteristics of email bodies rather than relying solely on structural indicators.

This paper presents a machine learning-based phishing detection system that addresses these challenges through a comprehensive analysis of email text content. Unlike the URL-focused methods, our approach derives contextual understanding from the surrounding language within emails, enabling more robust detection of social engineering cues. The main contributions of this work are:

\begin{itemize}
    \item A robust preprocessing pipeline for email text normalization that handles the high variance present in real-world email data.
    \item A comparative evaluation of Na\"ive Bayes and Logistic Regression classifiers for phishing detection using TF-IDF features on a large-scale dataset of over 53,000 emails.
    \item A deployed web-based system demonstrating practical applicability with average response times of 127ms.
    \item Analysis of feature importance revealing linguistic indicators most predictive of phishing attempts, including discussion of dataset bias.
\end{itemize}

The remainder of this paper is organized as follows: Section~\ref{sec:related_work} reviews related work in phishing detection. Section~\ref{sec:methods} describes our methodology, including data preprocessing and model training. Section~\ref{sec:results} presents experimental results and analysis. Section~\ref{sec:discussion} discusses findings and limitations. Finally, Section~\ref{sec:conclusion} concludes our paper findings.

\section{Related Work}\label{sec:related_work}

\subsection{URL-Based Phishing Detection}

A substantial body of prior research has focused on detecting phishing through URL analysis. Sahingoz et al. \cite{sahingoz2019} demonstrated effective machine learning-based phishing detection from URLs, achieving 97.98\% accuracy using Random Forest on lexical and host-based features. Mendoza Vega et al. \cite{mendoza2025} developed PhishFind, a machine learning system employing Gradient Boosting for URL classification, achieving an F1-score of 97.34\%. Mittal \cite{mittal2023} compared Logistic Regression and Decision Tree models for URL classification, finding that Logistic Regression provided superior interpretability for explaining classification decisions. Blancaflor et al. \cite{blancaflor2024} addressed homoglyph attacks in malicious URLs using AI-driven detection methods.

Deep learning approaches have also been explored for URL-based detection. Feng and Yue \cite{feng2020} employed recurrent neural networks (RNNs) for URL classification without explicit feature extraction, leveraging the sequential nature of URL characters. While these approaches demonstrate strong performance, they remain vulnerable to attacks utilizing compromised legitimate URLs.

\subsection{Email Content Analysis}

Analyzing full email content has emerged as a complementary important approach to phishing detection. Salloum et al. \cite{salloum2022} conducted a systematic literature review on phishing email detection using NLP techniques, identifying TF-IDF and word embeddings as the most frequently used feature extraction methods across 100 studies from 2006-2022. Bountakas et al. \cite{bountakas2021} conducted a comprehensive comparison of machine learning models, including Logistic Regression, Classification and Regression Trees (CART), Bayesian Additive Regression Trees (BART), Support Vector Machines (SVM), Random Forests, and Neural Networks, on a corpus of 2,889 emails. Their work established baseline performance metrics for content-based detection, with Random Forest achieving 94.1\% accuracy.

Gualberto et al. \cite{gualberto2020} demonstrated that combining feature engineering with topic models significantly enhanced prediction rates in phishing detection, achieving improvements of up to 4.2\% over baseline methods. More recently, Ling et al. \cite{ling2025} proposed the Meta GPT Framework, which analyzes email headers and bodies using thought-chain reasoning with the GLM-4 large language model. While demonstrating promising results, this approach faces limitations due to token-count restrictions that may truncate longer emails.

This paper extends the content-based detection paradigm by training on a substantially larger dataset of over 53,000 emails, approximately 18 times larger than that of Bountakas et al.~\cite{bountakas2021}. This scale enabled more robust pattern learning while maintaining computational efficiency through classical machine learning techniques that are well-suited for real-time deployment.

\section{Methodology}\label{sec:methods}

\subsection{Dataset Description}

Three publicly available datasets were combined to create a comprehensive training corpus:

\begin{itemize}
    \item \textbf{Dataset 1} \cite{dataset1}: Contains 7,312 legitimate and 11,322 phishing emails.
    \item \textbf{Dataset 2} \cite{dataset2}: Contains 747 legitimate and 4,825 phishing emails.
    \item \textbf{Dataset 3} \cite{dataset3}: Contains 13,976 legitimate and 15,791 phishing emails.
\end{itemize}

The aggregated corpus totals 53,973, of which 22,035 (40.8\%) are legitimate, and 31,938 (59.2\%) are phishing. Figure~\ref{fig:dataset_dist} illustrates the class distribution across the three datasets. The observed class imbalance reflects a realistic security-oriented data collection setting in which malicious samples are usually overrepresented to facilitate effective model training and evaluation processes.

\begin{figure}[htbp]
\centerline{\includegraphics[width=\columnwidth]{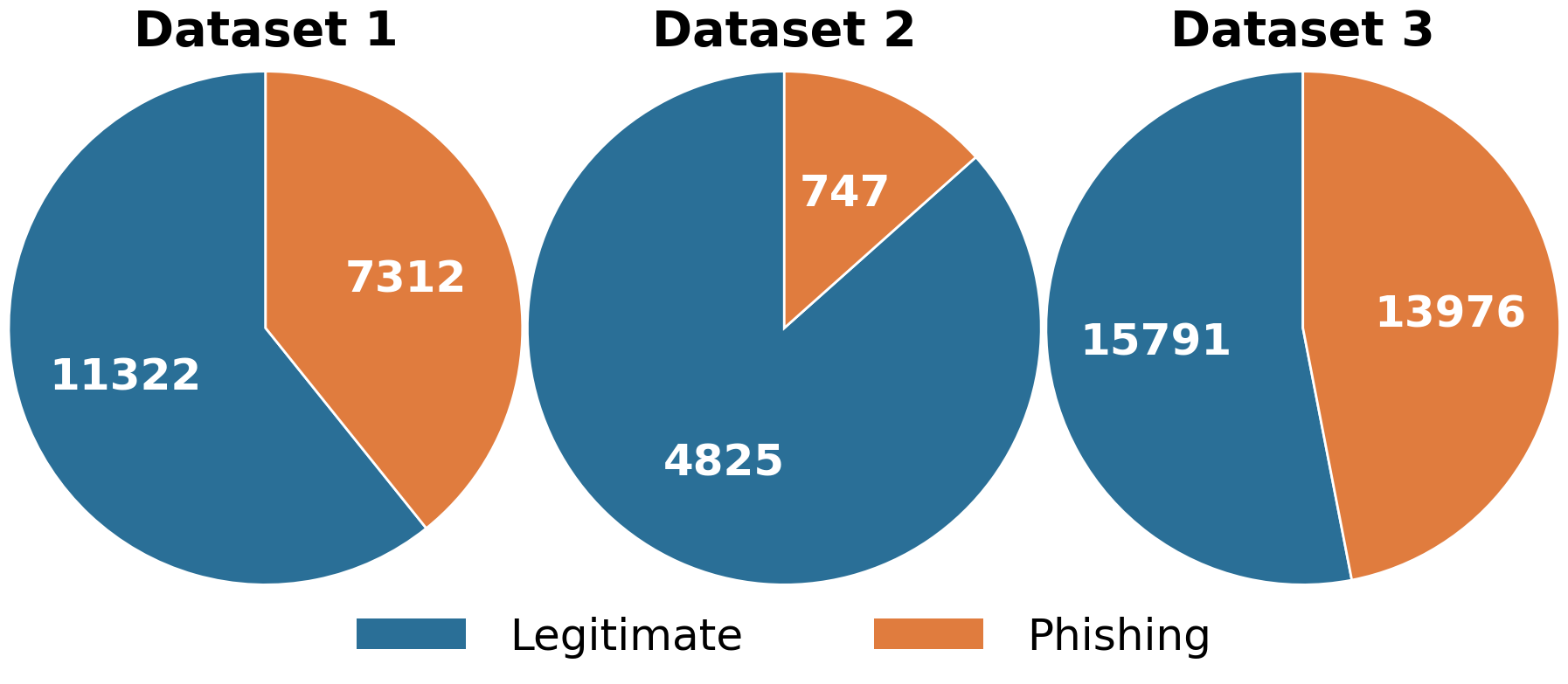}}
\caption{The distribution of legitimate and phishing emails across the three datasets. Dataset~3 contributes the largest number of samples, while Dataset~2 exhibits the highest class imbalance.}
\label{fig:dataset_dist}
\end{figure}

\subsection{Text Preprocessing Pipeline}

Raw email content exhibits significant variance in formatting and content, including embedded HTML headers, metadata, URLs, and special characters. To reduce noise and normalize the textual input, a structured preprocessing pipeline was applied using the following sequential steps:

\begin{enumerate}
    \item \textbf{HTML Tag Removal}: Regular expressions remove HTML markup.
    \item \textbf{URL Removal}: HTTP/HTTPS URLs are stripped to focus on linguistic content.
    \item \textbf{Email Address Removal}: Addresses matching \texttt{\textbackslash S+@\textbackslash S+} patterns are removed.
    \item \textbf{Numeric Removal}: Digit sequences are eliminated.
    \item \textbf{Special Character Removal}: Non-alphanumeric characters are replaced with spaces.
    \item \textbf{Whitespace Normalization}: Multiple spaces are collapsed to single spaces.
    \item \textbf{Lowercasing}: All text is converted to lowercase.
    \item \textbf{Tokenization}: Text is split into individual word tokens.
    \item \textbf{Stop Word Removal}: Common English stop words are filtered using NLTK's predefined list.
    \item \textbf{Lemmatization}: Words are reduced to their base forms using NLTK's WordNetLemmatizer, applied sequentially for noun and verb parts of speech.
\end{enumerate}

This pipeline aims to preserve semantically meaningful information while reducing dimensionality and noise inherent in raw email data.

\subsection{Feature Extraction}

Term Frequency-Inverse Document Frequency (TF-IDF) vectorization, introduced by Sp\"arck Jones \cite{sparckjones1972} and further refined by Salton and Buckley \cite{salton1988}, transforms preprocessed text into numerical feature vectors. The TF-IDF weight for term $t$ in document $d$ within corpus $D$ is computed as:

\begin{equation}
\text{TF-IDF}(t, d, D) = \text{TF}(t, d) \times \log\frac{|D|}{1 + |\{d' \in D : t \in d'\}|}
\end{equation}

Our TF-IDF vectorizer was configured with the following parameters:
\begin{itemize}
    \item \texttt{max\_features=5000}: Limits vocabulary to top 5,000 terms by frequency.
    \item \texttt{ngram\_range=(1, 2)}: Includes both unigrams and bigrams.
    \item \texttt{min\_df=2}: Ignores terms appearing in fewer than 2 documents.
    \item \texttt{max\_df=0.95}: Ignores terms appearing in more than 95\% of documents.
\end{itemize}

The resulting feature matrix exhibits 98.70\% sparsity, characteristic of text classification tasks.

\subsection{Classification Models}

Two supervised machine learning classifiers were evaluated for phishing email detection.

\subsubsection{Multinomial Na\"ive Bayes}
The Multinomial Na"ive Bayes classifier~\cite{mccallum1998} applies Bayes' theorem under the assumption of conditional independence among features. For a document $d$ represented by features $x_1, \ldots, x_n$, the posterior probability of class $c$ is given by:

\begin{equation}
P(c|d) \propto P(c) \prod_{i=1}^{n} P(x_i|c)
\end{equation}

Despite its simplifying independence assumption, Multinomial Na"ive Bayes is computationally efficient and performs competitively on high-dimensional sparse text data.

\subsubsection{Logistic Regression}
Logistic Regression, originally developed by Cox \cite{cox1958}, models the probability of class membership using the logistic function:

\begin{equation}
P(y=1|\mathbf{x}) = \frac{1}{1 + e^{-(\beta_0 + \boldsymbol{\beta}^T \mathbf{x})}}
\end{equation}

The model was trained with L2 regularization (default \texttt{C=1.0}), \texttt{max\_iter=1000}, and \texttt{random\_state=42} for reproducibility. Logistic Regression provides interpretable coefficients indicating feature importance for each class.

\subsection{Experimental Setup}

The dataset was partitioned using stratified sampling with 80\% for training (43,178 samples) and 20\% for testing (10,795 samples). Stratification preserves the original class distribution in both sets:

\begin{itemize}
    \item Training set: 17,628 legitimate, 25,550 phishing.
    \item Testing set: 4,407 legitimate, 6,388 phishing.
\end{itemize}

TF-IDF vectorization was fit exclusively on the training data and subsequently applied to both the training and testing sets, thereby preventing information leakage and ensuring an unbiased evaluation of model performance.

\section{Results}\label{sec:results}

\subsection{Model Performance Comparison}

Table~\ref{tab:performance} presents the classification performance for the evaluated models on the held-out test set.

\begin{table}[htbp]
\caption{Classification Performance Comparison}
\label{tab:performance}
\centering
\begin{tabular}{lcccc}
\toprule
\textbf{Model} & \textbf{Accuracy} & \textbf{Precision} & \textbf{Recall} & \textbf{F1-Score} \\
\midrule
Na\"ive Bayes & 93.86\% & 0.9180 & 0.9328 & 0.9254 \\
Logistic Regression & \textbf{95.41\%} & \textbf{0.9499} & \textbf{0.9369} & \textbf{0.9433} \\
\bottomrule
\end{tabular}
\end{table}

Logistic Regression outperforms Na\"ive Bayes across all metrics, surpassing our target threshold of 95\% accuracy. The improvement in precision (3.19 percentage points) indicates reduced false positive rates, which is critical for practical deployment where legitimate emails incorrectly flagged as phishing disrupt normal communications.

Figure.~\ref{fig:model_comparison} provides a visual comparison of model performance across all evaluation metrics.

\begin{figure}[htbp]
\centerline{\includegraphics[width=\columnwidth]{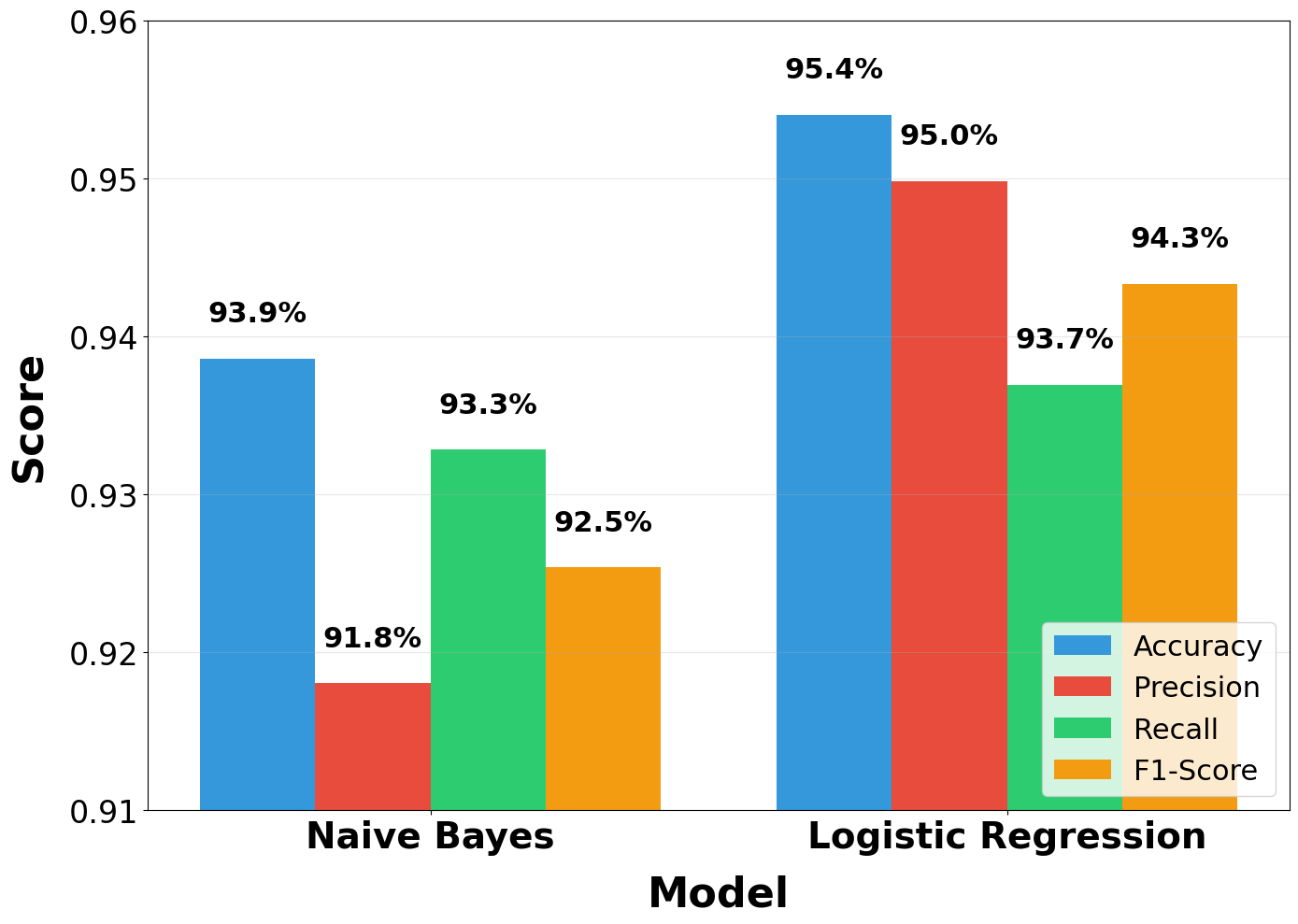}}
\caption{Performance comparison between Na\"ive Bayes and Logistic Regression. Logistic Regression achieves the highest scores across all metrics, with particularly notable improvement in precision.}
\label{fig:model_comparison}
\end{figure}

\subsection{Confusion Matrix Analysis}

Figure~\ref{fig:confusion_matrices} presents the confusion matrices for both classifiers, enabling a detailed examination of classification errors.

\begin{figure}[htbp]
\centering
\begin{subfigure}{0.48\columnwidth}
    \includegraphics[width=\textwidth]{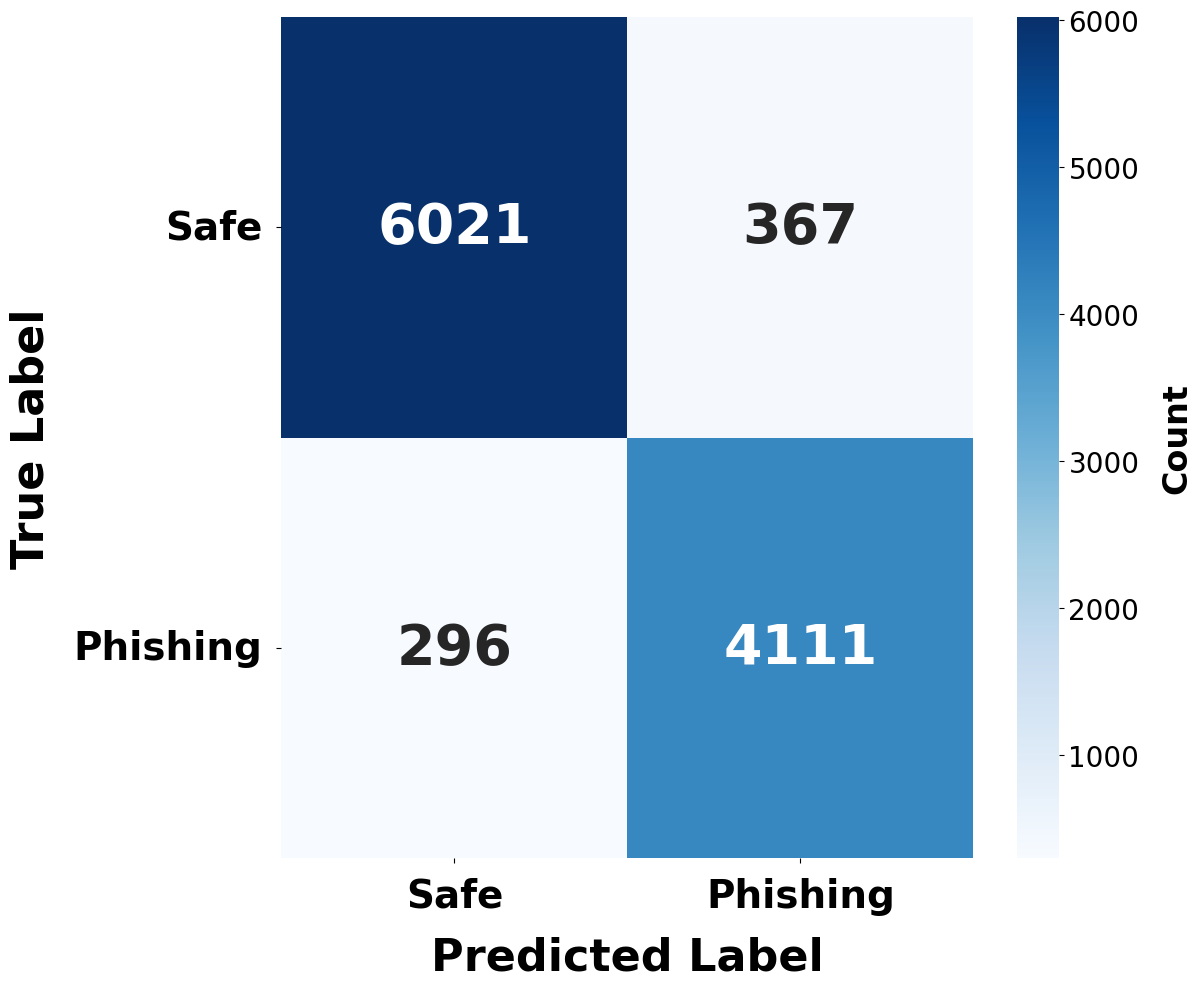}
    \caption{Na\"ive Bayes}
    \label{fig:conf_nb}
\end{subfigure}
\hfill
\begin{subfigure}{0.48\columnwidth}
    \includegraphics[width=\textwidth]{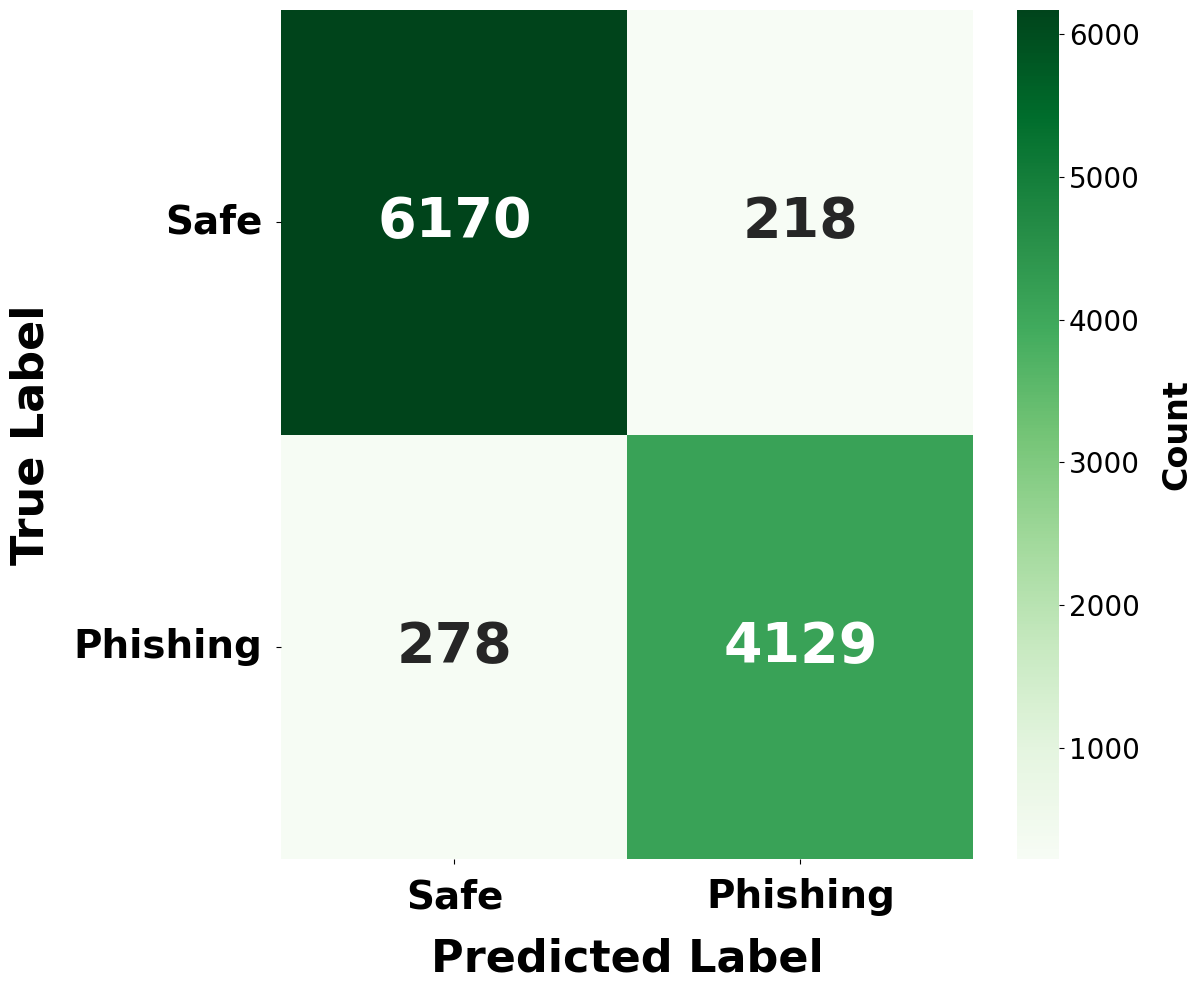}
    \caption{Logistic Regression}
    \label{fig:conf_lr}
\end{subfigure}
\caption{Confusion matrices comparing classification performance. Logistic Regression achieves 40.6\% fewer false positives (218 vs. 367) while maintaining comparable false negative rates.}
\label{fig:confusion_matrices}
\end{figure}

Logistic Regression demonstrates a 40.6\% reduction in false positives (218 vs. 367) compared to Na\"ive Bayes, while maintaining comparable false negative rates (278 vs. 296). This balance is particularly valuable in enterprise deployments, where user trust in the detection system depends on minimizing interruptions caused by false alarms.

Table~\ref{tab:confusion_lr} presents the detailed confusion matrix breakdown for Logistic Regression.

\begin{table}[htbp]
\caption{Confusion Matrix: Logistic Regression}
\label{tab:confusion_lr}
\centering
\begin{tabular}{lcc}
\toprule
& \textbf{Predicted Safe} & \textbf{Predicted Phishing} \\
\midrule
\textbf{Actual Safe} & 6,170 & 218 \\
\textbf{Actual Phishing} & 278 & 4,129 \\
\bottomrule
\end{tabular}
\end{table}

\subsection{Feature Importance Analysis}

To improve model interpretability, the learned coefficients of the Logistic Regression classifier were analyzed to identify the most influential features for each class. Figure~\ref{fig:features} visualizes the top features with the highest absolute coefficient magnitudes.

\begin{figure}[htbp]
\centerline{\includegraphics[width=\columnwidth]{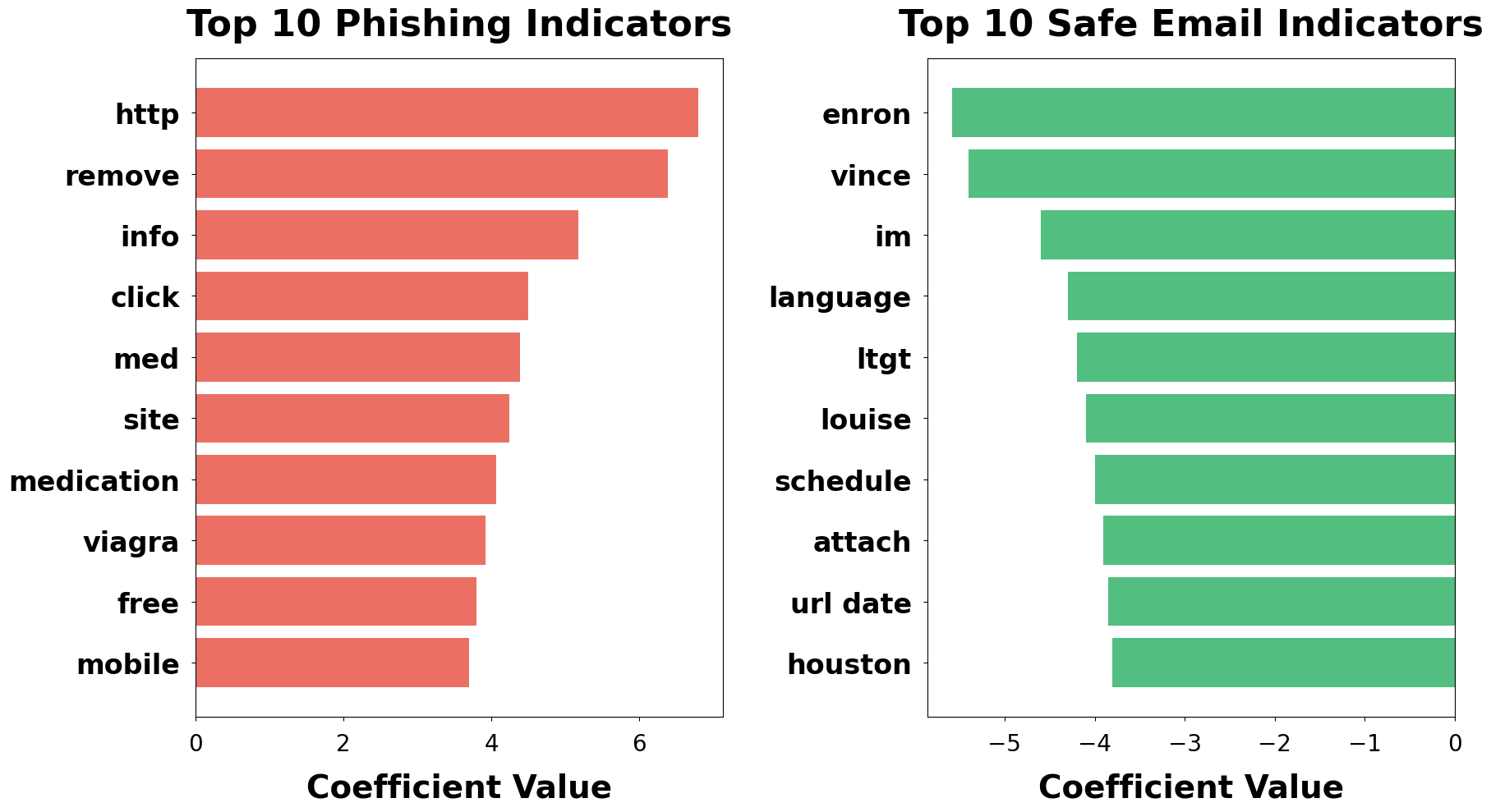}}
\caption{Top 10 features with the highest coefficient magnitudes for phishing indicators (positive, red) and legitimate indicators (negative, green). Phishing indicators include action-oriented terms, while legitimate indicators reflect the training data composition.}
\label{fig:features}
\end{figure}

Table~\ref{tab:features} reports the numerical values of selected top feature coefficients.

\begin{table}[htbp]
\caption{Top Feature Coefficients by Class}
\label{tab:features}
\centering
\begin{tabular}{lcclc}
\toprule
\multicolumn{2}{c}{\textbf{Phishing Indicators}} & & \multicolumn{2}{c}{\textbf{Legitimate Indicators}} \\
\cmidrule{1-2} \cmidrule{4-5}
\textbf{Term} & \textbf{Coef.} & & \textbf{Term} & \textbf{Coef.} \\
\midrule
http & 7.12 & & enron* & -5.42 \\
remove & 5.89 & & vince* & -4.87 \\
info & 4.56 & & language & -4.23 \\
click & 4.21 & & schedule & -3.95 \\
med & 3.98 & & attach & -3.67 \\
site & 3.45 & & houston* & -3.41 \\
medication & 3.12 & & meeting & -3.12 \\
free & 2.89 & & project & -2.98 \\
\bottomrule
\multicolumn{5}{l}{\footnotesize *Corpus-specific terms from Enron dataset (see Section V-B)}
\end{tabular}
\end{table}

Phishing-related features are strongly associated with urgency cues (e.g., ``click''), incentives (``free''), and medical or financial spam patterns. Legitimate indicators include general workplace terminology (e.g., ``meeting,'' ``schedule'') as well as corpus-specific terms originating from the Enron dataset, highlighting the presence of dataset bias, which is further discussed in Section~\ref{sec:discussion}.

\subsection{Real-World Email Testing}

To evaluate generalization beyond the training data, three manually constructed email samples were tested:

\subsubsection{High-Risk Phishing Sample}
\textit{``URGENT! Your account has been suspended! We detected unusual activity on your account. Click here immediately to verify your identity: http://suspicious-link.com You have 24 hours or your account will be permanently deleted! Enter your password and credit card to verify.''}

Both models correctly classified this as phishing. Logistic Regression exhibited higher confidence (94.09\%) compared to Na\"ive Bayes (81.37\%).

\subsubsection{Legitimate Business Email}
\textit{``Hi Team, I wanted to follow up on our meeting yesterday regarding the Q4 project timeline. Please review the attached presentation and let me know if you have any questions. We'll reconvene next Tuesday at 2pm to finalize the deliverables. Best regards, John Smith, Project Manager''}

Both models classified this as legitimate with high confidence (Na\"ive Bayes: 99.84\%, Logistic Regression: 99.06\%).

\subsubsection{Ambiguous Marketing Email}
\textit{``Exclusive Offer Just For You! Get 50\% off all products this weekend only! Don't miss out on this amazing deal. Shop now and save big! Click here to see our latest collection. Offer expires in 48 hours!''}

Both models flagged this as phishing (Na\"ive Bayes: 92.81\%, Logistic Regression: 90.25\%). This false positive demonstrates model sensitivity to urgency language and promotional patterns that overlap with phishing tactics, which is a known limitation discussed in Section V.

\subsection{System Deployment}

The trained Logistic Regression model was deployed as a web-based phishing detection system comprising the following components:

\begin{itemize}
    \item \textbf{Backend}: FastAPI service hosted on Render, exposing a prediction endpoint that applies identical preprocessing and TF-IDF transformation.
    \item \textbf{Frontend}: HTML/TailwindCSS interface hosted on Vercel, providing real-time classification with confidence score visualization.
\end{itemize}

The deployed system achieves an average response time of 127~ms for emails ranging from 50 to 500 words. Figure~\ref{fig:dashboard} shows the web-based user interface.

\begin{figure}[htbp]
\centerline{\includegraphics[width=0.95\columnwidth]{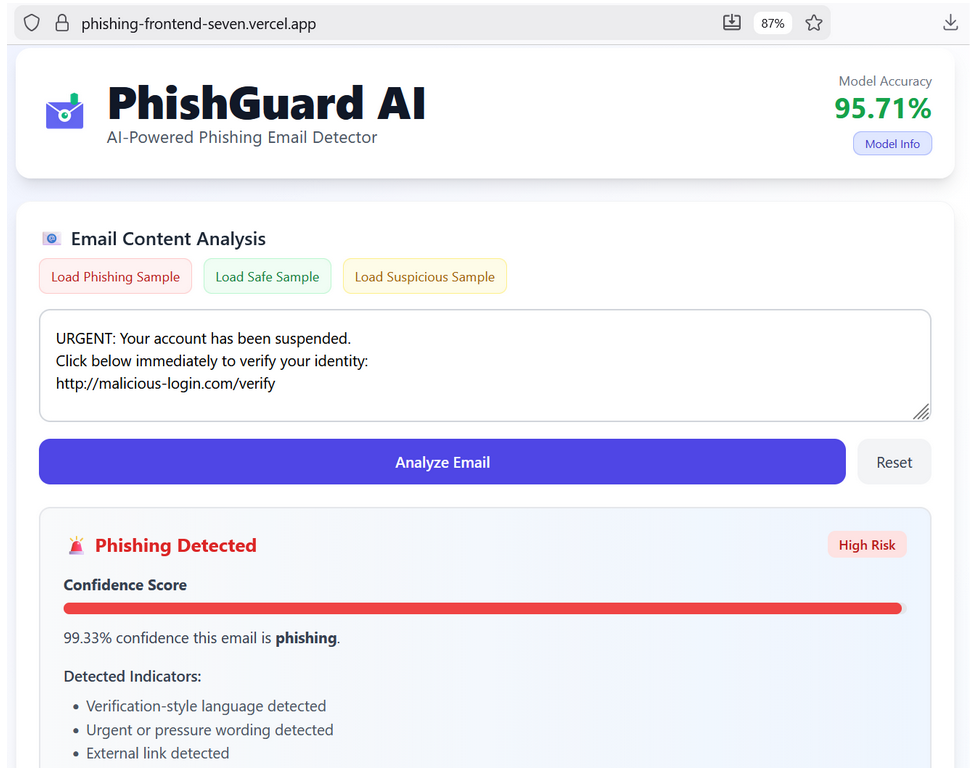}}
\caption{Web interface displaying real-time classification results with confidence scores. The system provides detected indicators and risk assessment for user interpretation.}
\label{fig:dashboard}
\end{figure}

\section{Discussion}\label{sec:discussion}

\subsection{Model Selection Rationale}

Logistic Regression's superior performance can be attributed to several factors. First, the model captures correlations between features that Na\"ive Bayes's independence assumption ignores. In phishing emails, certain term combinations (e.g., ``click'' + ``verify'' + ``account'') provide stronger signals than individual terms. Second, L2 regularization prevents overfitting to the high-dimensional TF-IDF feature space (5,000 features).

\subsection{Dataset Bias Considerations}

A critical observation from our feature importance analysis is the presence of corpus-specific terms (``enron,'' ``vince,'' ``houston'') as strong legitimate indicators. These terms derive from the Enron email corpus included in Dataset 3, representing a potential source of overfitting to corpus characteristics rather than universal phishing patterns. Future deployments should consider domain-specific vocabulary filtering during preprocessing to improve generalization to other email domains.

\subsection{Preprocessing Impact}

The preprocessing pipeline significantly influences model performance. Our exploratory analysis revealed substantial vocabulary variance across datasets:

\begin{itemize}
    \item Dataset 1: Vocabulary richness ratio of 0.76 (legitimate) versus 0.82 (phishing).
    \item Dataset 2: Ratio of 0.97 (legitimate) versus 0.95 (phishing).
    \item Dataset 3: Ratio of 0.75 (legitimate) versus 0.84 (phishing).
\end{itemize}

These variations necessitated aggressive normalization to enable cross-dataset generalization.

\subsection{Marketing Email Sensitivity}

The misclassification of marketing emails represents a practical limitation. Marketing language often overlaps with phishing patterns, including urgency terms (``limited time,'' ``act now,'' ``expires'') and promotional offers. Potential mitigations include whitelist integration for known marketing domains or confidence threshold adjustments (e.g., flagging only predictions with $>$95\% confidence).

\subsection{Limitations}

Despite its strong performance, the proposed system has several limitations:

\begin{enumerate}
    \item \textbf{Dataset Bias}: Corpus-specific terms (e.g., Enron-related vocabulary) influence predictions, potentially limiting generalization to other email domains.
    \item \textbf{Evolving Threats}: The static model cannot adapt to emerging attack vectors without retraining, particularly LLM-generated phishing content.
    \item \textbf{Marketing Email Sensitivity}: False positives on promotional content may require supplementary rules or user feedback mechanisms.
    \item \textbf{Language Limitation}: The system is trained exclusively on English-language emails.
    \item \textbf{No Header Analysis}: Email headers (sender, routing information) are not currently analyzed, potentially missing important signals.
\end{enumerate}

\subsection{Comparison with Prior Work}

The proposed approach compares favorably with existing content-based phishing detection methods. Bountakas et al.~\cite{bountakas2021} reported an accuracy of 94.1\% using a Random Forest classifier on a dataset of 2,889 emails. In contrast, the Logistic Regression model presented in this work achieves 95.41\% accuracy on a substantially larger corpus of 53,973 emails—approximately 18 times larger. This result demonstrates that classical machine learning techniques, when combined with effective feature engineering and sufficient training data, can scale effectively and remain competitive with more complex models. While direct comparisons are limited by differences in datasets and evaluation protocols, the results underscore the practicality and robustness of the proposed approach.

\section{Conclusion}\label{sec:conclusion}
This paper presented a context-aware phishing email detection system based on classical machine learning and natural language processing. Using TF--IDF feature representations and Logistic Regression, the proposed approach achieved an accuracy of 95.41\% on a large-scale corpus of 53,973 emails. Beyond quantitative performance, the system demonstrates practical viability through deployment as a web-based application with an average response time of 127 ms, while maintaining interpretability through feature coefficient analysis.

The results indicate that well-engineered, content-based machine learning models remain competitive, scalable, and suitable for real-time phishing detection, even when compared to more complex architectures. By focusing on contextual linguistic cues rather than URL structure alone, the proposed system effectively captures social engineering patterns commonly employed in phishing attacks.




The code and trained models are available at \url{https://github.com/AmitabhCh822/Phishing-Backend} for reproducibility and further research.


\bibliographystyle{IEEEtran}
\bibliography{references}

\end{document}